\documentclass[article,amsmath,amssymb]{revtex4-2}
\usepackage{graphicx}
\usepackage{dcolumn}
\usepackage[colorlinks]{hyperref}
\usepackage{bm}

\begin{document}

\title{Mean-Field Description of Cooperative Scattering by Atomic Clouds}

\author{Nicola Piovella}

\affiliation{Dipartimento di Fisica "Aldo Pontremoli", Universit\`{a} degli Studi di Milano, Via Celoria 16, I-20133 Milano, Italy}

\begin{abstract}
We present analytic expressions for the scattering of light by an extended atomic cloud. We obtain the solution for the mean-field excitation of different atomic spherical distributions driven by an uniform laser, including the initial build-up, the steady-state and the decay after the laser is switched off. We show that the mean-field model does not describe subradiant scattering, due to negative interference of the photons scattered by $N$ discrete atoms.
\end{abstract}

\maketitle

\section{Introduction}

The cooperative emission from a system of $N$ two-level excited atoms has been the object of an intense investigation is the past, starting from the pioneering studies on Dicke superradiance \cite{Dicke1954}. On the other hand, the diffusive regime of cooperative scattering in dense medium has been study extensively in the past by a diagramatic approach \cite{hayden2001coherent}, where  light travels over a distance much larger than the mean free path. More recently, optical properties of atomic clouds beyond the single-atom level have been studied theoretically \cite{Cherroret2016,Kwong2019} and experimentally \cite{Beugnon2018}, where the connections between the perturbative diffusive theory and the coupled-dipole description have been investigated. Also, cooperative scattering from dense cold atomic clouds has been the object of intense investigation \cite{Browaeys2016}.
These studies are complementary to a different regime where the light scattering induces a dipole-dipole interaction between the atom pairs, leading to the cooperative processes of superradiance and subradiance. This regime is characterized to be dominated by single-scattering of photons by many atoms, whereas to the diffusive regime is dominated by multiple scattering. The transition between single and multiple scattering is controlled by the optical thickness parameter $b(\Delta)=b_0/(1+4\Delta^2/\Gamma^2)$\cite{labeyrie2003slow,guerin2017light}, where $b_0$ is the resonant optical thickness, $\Delta$ is the detuning of the laser frequency from the atomic resonance frequency and $\Gamma$ is the transition linewidth. In this contest, a new kind of single-photon superradiance has been proposed by Scully and coworkers \cite{Scully2006,Svidzinsky2008,Svidzinsky2010}, from an extended ensemble of $N$ atoms prepared by absorption of a single photon and exhibiting superradiant decay. A bridge between this single-photon superradiance and the more classical process of cooperative scattering of an incident laser by $N$ atoms \cite{Lehmberg1970} has been proposed by a series of theoretical and experimental papers \cite{Courteille2010,Bienaime2010,Bienaime2013b,Chabe2014,Bachelard2016}. A more intriguing effect in such systems is subradiance, also initially proposed by Dicke in 1954 \cite{Dicke1954}, i.e. destructive interference effect leading to the partial trapping of light in the system. This effect has been predicted \cite{Bienaime2012} and then observed \cite{Guerin2016} in a system of driven cold atoms, after that the laser is abruptly switched off and the emitted photons detected in a given direction. From a theoretical point of view, subradiance has been investigated mostly studying the eigenvalues of the system and identified in the most long-lived modes, surviving after the more fast superradiant modes have been extinguished \cite{Bellando2014,Guerin2017,Cottier2018}. The analysis was based on the numerical solution of the coupled-dipole model of $N$ two-level atoms driven by an uniform laser field. A continuous-distribution version of this model allows for an analytical treatment of the problem. This has been done extensively in a series of papers by Svidzinsky et al. \cite{Svidzinsky2008,Svidzinsky2010}, considering the temporal decay of the system initially prepared in some given excited state. The stationary problem of the system driven by an uniform laser  has been studied in terms of the collective modes in ref.\cite{Bachelard2011,Bachelard2012}. However, the cooperative decay after that the laser has been switched off has not yet been studied by using the continuous-distribution model (or the so-called mean-field (MF) model), except in ref.\cite{Cottier2018} where however the solution has been obtained numerically.

The aim of this paper is to provide analytical expressions for the excitation of the driven system and for the scattered light intensity. This will encompass both the analytical work by Svidzinsky et al. \cite{Svidzinsky2008,Svidzinsky2010}, who did not consider a driven system, and the numerical results of ref.\cite{Cottier2018}. 

We outline that the MF model assumes a coherent interaction between the scatterers, neglecting granularity and fluctuations in the atomic distribution. These ingredients are necessary in order to describe the random walk of the scattered photons, leading to the diffusive regime for sufficiently dense samples \cite{labeyrie2003slow,guerin2017light}. Hence, the MF model is not able to describe the diffusive regime, where the atom scatters a photon many times within a mean-free path. As already mentioned, multiple-scattering regime is characterized by a large optical thickness $b=L/\ell_{sc}$, where $L$ is the size of the medium, $\ell_{sc}=1/n\sigma_{sc}$ is the mean-free pass, $n$ is the atomic density, $\sigma_{sc}=(6\pi/k^2)/(1+4\Delta^2/\Gamma^2)$ is the scattering cross-section and $k$ is the laser wavenumber. Hence, the MF model is valid for small optical thickness $b(\Delta)=b_0/(1+4\Delta^2/\Gamma)$, i.e. for large resonant optical thickness $b_0$ and large detuning, such that $b(\Delta)\ll 1$.

The paper is organized as follows. In sec. \ref{sec:2} we present the general MF equations for a continuous atomic distribution. In sec. \ref{sec:3} we derive the expression for the average quantities and the scattered light intensity and power. The particular cases of uniform, parabolic and Gaussian radial distribution are discussed in sec. \ref{sec:4} and compared with the numerical solution of the discrete model. Conclusions are summarized in sec. \ref{sec:5}.

\section{General equations}\label{sec:2}

From a microscopic point of view and using a dipole approximation, our medium is composed of an ensemble of $N$
two-level atoms with position $\mathbf{r}_j$, whose atomic transition has frequency $\omega_a$, linewidth $\Gamma$ and dipole $d$ (polarization effects are neglected). The system is  driven by a monochromatic plane wave with electric field $E_0$, frequency $\omega_0$ and  wave vector $\mathbf{k}_0$, detuned from the atomic transition by $\Delta_0=\omega_0-\omega_a$. In the linear regime and in the Markov approximation (valid if the decay time is larger than the photon time-of-flight through the atomic cloud), the problem reduces to the following differential equation for the atomic dipole amplitudes $\beta_j$ \cite{Bienaime2011}:
\begin{eqnarray}
  \frac{d\beta_j}{dt} &=&
  \left(i\Delta_0-\frac{\Gamma}{2}\right)\beta_j-\frac{i\Omega_0}{2}e^{i\mathbf{k}_0\cdot\mathbf{r}_j}-\frac{\Gamma}{2}
  \sum_{m\neq j}G_{jm}
  \beta_m(t)\label{betaj}.
\end{eqnarray}
where $\Omega_0=dE_0/\hbar$ is the Rabi frequency and
\begin{equation}
G_{jm}=\frac{\exp(ik_0|\mathbf{r}_j-\mathbf{r}_m|)}{ik_0|\mathbf{r}_j-\mathbf{r}_m|}=\frac{\sin(k_0|\mathbf{r}_j-\mathbf{r}_m|)}{k_0|\mathbf{r}_j-\mathbf{r}_m|}-i\frac{\cos(k_0|\mathbf{r}_j-\mathbf{r}_m|)}{k_0|\mathbf{r}_j-\mathbf{r}_m|}
\end{equation}
The kernel $G_{jm}$ describes the coupling between the dipoles, mediated by the photons exchanged between the dipoles. It has a real component (sine term), describing the cooperative atomic decay, and an imaginary component (cosine term) describing the cooperative Lamb shift \cite{Scully2009}. The latter
becomes significant when the number of atoms in a cubic optical wavelength, $n\lambda^3$, is larger than unity, such that
the contribution from the virtual photons becomes relevant.

In light scattering experiments, disorder plays a role when the number of atoms projected onto a cross section perpendicular to the incident beam is small enough so that a light mode focused down to the diffraction limit (that is $\lambda^2$) would be able to resolve and count the atoms. In
other words, the stochastic fluctuations induced by the random positions of the atoms can be neglected when the total number of atoms $N$ is larger than the number of modes $\sigma^2$ (where $\sigma=k_0 R$ and $R$ is the transverse size of the system) that fit
into the cloud’s cross section, i.e. when the optical density is $b_0=3N/\sigma^2\gg 1$. Under this hypothesis, the particles can be
described by a smooth density $n(\mathbf{r})$ and their probability to be excited by a field $\beta(\mathbf{r},t)$. 
By approximating the sum over $j$ by an integral over the smooth density, i.e. $\sum_j\rightarrow
\int d\mathbf{r}n(\mathbf{r})$, Eq.(\ref{betaj}) turns into
\begin{align}
    \frac{\partial\beta(\mathbf{r},t)}{\partial t}= \left(i\Delta_0-\frac{\Gamma}{2}\right)\beta(\mathbf{r},t)
    - \frac{i}{2}\Omega_0 e^{i\mathbf{k}_0\cdot\mathbf{r}}-\frac{\Gamma}{2}
    \int d\mathbf{r}' n(\mathbf{r}')\frac{\exp(ik_0|\mathbf{r}-\mathbf{r}'|)}{ik_0|\mathbf{r}-\mathbf{r}'|}\beta(\mathbf{r}',t).\label{beta}
\end{align}
Using
\begin{equation}\label{sinc}
    \frac{\exp(ik_0|\mathbf{r}-\mathbf{r}'|)}{ik_0|\mathbf{r}-\mathbf{r}'|}=
    4\pi\sum_{n=0}^\infty\sum_{m=-n}^{n}j_n(k_0 r_<)Y_{n,m}(\theta,\phi)Y^*_{n,m}(\theta',\phi')h^{(1)}_n(k_0 r_>)
\end{equation}
where $Y_{n,m}(\theta,\phi)$ are the spherical harmonics, $j_n(r)$ and $h^{(1)}_n(r)=j_n(r)+iy_n(r)$ are the spherical Bessel and Hankel functions of first kind, respectively, and 
$r_<$ ($r_>$) is the smaller (larger) between $r$ and $r'$. Taking $\theta$ as the polar angle with respect the direction of the  wave vector $\mathbf{k}_0$,
we can expand
\begin{equation}\label{betaexp}
    \beta(\mathbf{r},t)=\sum_{n,m}\alpha_{n,m}(t)\beta_{n}(r)Y_{n,m}(\theta,\phi).
\end{equation}
By substituting it in Eq.(\ref{beta}) and assuming a radial distribution, $n(r)$, we obtain
\begin{eqnarray}
    &&\sum_{n',m'}\left\{
    \dot\alpha_{n',m'}-\left(i\Delta_0-\frac{\Gamma}{2}\right)\alpha_{n',m'}\right\}
    \beta_{n'}(r)Y_{n',m'}(\theta,\phi)=-\frac{i}{2}\Omega_0
    e^{ik_0r\cos\theta}\nonumber\\
    &-&\frac{\Gamma}{2}(4\pi)
    \int_0^\infty dr' r'^2 n(r')\int d\Omega'
    \sum_{n'',m''}j_{n''}(k_0 r_<)Y_{n'',m''}(\theta,\phi)Y^*_{n'',m''}(\theta',\phi')h^{(1)}_{n''}(k_0 r_>)\nonumber\\
    &\times &
    \sum_{n',m'}\alpha_{n',m'}\beta_{n'}(r')Y_{n',m'}(\theta',\phi'),
    \label{Ebeta}
\end{eqnarray}
where $d\Omega'=d\phi'\sin\theta'd\theta'$.
Since
\begin{equation}
\int_0^{2\pi}d\phi\int_0^\pi d\theta\sin\theta\,
Y^*_{n,m}(\theta,\phi)Y_{n',m'}(\theta,\phi)=\delta_{n,n'}\delta_{m,m'}\label{norm}
\end{equation}
and
\begin{eqnarray}
    \int_0^{2\pi}d\phi\int_0^\pi d\theta\sin\theta\,
    Y^*_{n,m}(\theta,\phi)
    e^{ik_0r\cos\theta}=2\delta_{m,0}\sqrt{\pi(2n+1)}i^n
    j_n(k_0 r),\label{int}
\end{eqnarray}
multiplying Eq.(\ref{Ebeta}) by $Y_{n,m}^*(\theta,\phi)$ and integrating over the angles, we obtain:
\begin{eqnarray}
    \left\{
    \dot\alpha_{n,m}-\left(i\Delta_0-\frac{\Gamma}{2}\right)\alpha_{n,m}\right\}\beta_n(r)
    &=&-i\Omega_0\delta_{m,0}\sqrt{\pi(2n+1)}i^n
    j_n(k_0 r)\nonumber\\
    &-&\frac{\Gamma}{2}(4\pi)\alpha_{n,m}
    \int_0^\infty dr' r'^2 n(r')j_{n}(k_0 r_<)h^{(1)}_{n}(k_0 r_>)\beta_n(r').\nonumber\\
    \label{Ebeta2}
\end{eqnarray}
If $\alpha_{n,m}(0)=0$, the only components different from zero are those for $m=0$. So, defining $\alpha_n=\sqrt{(2n+1)/4\pi}\,\alpha_{n,0}$ and since $Y_{n,0}(\theta,\phi)=\sqrt{(2n+1)/4\pi}\, P_n(\cos\theta)$ where $P_n(x)$ are the Legendre polynomial, we write
\begin{equation}\label{betaexp}
    \beta(r,\theta,t)=\sum_{n=0}^\infty\alpha_{n}(t)j_{n}(k_0r)P_n(\cos\theta)
\end{equation}
where $\alpha_n(t)$ is the solution of the following differential equation
\begin{eqnarray}
    \left\{
    \dot\alpha_{n}-\left(i\Delta_0-\frac{\Gamma}{2}\right)\alpha_{n}\right\}j_{n}(k_0r)
    &=&-i\frac{\Omega_0}{2}(2n+1)i^n
    j_n(k_0 r)-\frac{\Gamma}{2}F_n(r)\alpha_n  
    .\label{Ebeta3}
\end{eqnarray}
where
\begin{eqnarray}
F_n(r)&=&4\pi
\left\{h^{(1)}_{n}(k_0 r)\int_0^r dr' r'^2 n(r')j^2_{n}(k_0 r')
    +
    j_{n}(k_0 r)\int_r^\infty dr' r'^2 n(r')j_{n}(k_0r')h^{(1)}_{n}(k_0 r')\right\}\nonumber\\
    \label{Fn}
\end{eqnarray}
We observe that $F_n(r)$ has a real part and an imaginary part. The real part is $\mathrm{Re}\{F_n(r)\}=\lambda_n j_n(k_0r)$ where
\begin{equation}
\lambda_n=4\pi\int_0^\infty dr r^2 n(r)j^2_{n}(k_0 r)
\end{equation}
is the collective decay rate of the mode $n$
and it corresponds to the contribution of the sine term of the kernel of Eq.(\ref{beta}). The imaginary part is
\begin{equation}
\mathrm{Im}\{F_n(r)\}=4\pi\left\{y_{n}(k_0 r)\int_0^r dr' r'^2 n(r')j^2_{n}(k_0 r')
    +
    j_{n}(k_0 r)\int_r^\infty dr' r'^2 n(r')j_{n}(k_0r')y_{n}(k_0 r')\right\}
\end{equation}
and contributes to the cooperative Lamb shift, arising from the cosine term of the  kernel of Eq.(\ref{beta}). 
When the detuning $\Delta_0$ is much larger than the collective Lamb shift, the sine-kernel provides a good approximation to the solution.

\section{Average quantities}\label{sec:3}

Using the expansion (\ref{betaexp}), we can calculate the average:
\begin{eqnarray}
        \langle|\beta(t)|^2\rangle &=& \frac{2\pi}{N}\int_0^\pi
    d\theta\sin\theta\int_0^\infty 
    r^2n(r)|\beta(r,\theta,t)|^2dr\nonumber\\
    &=& \frac{2\pi}{N}\sum_{n,m=0}^\infty\alpha_m^*(t)\alpha_n(t)
    \int_0^\infty r^2n(r)j_{m}(k_0r)j_{n}(k_0r)dr \int_{-1}^1
    P_m(x)P_n(x)dx.\nonumber\\
\end{eqnarray}
Using 
\begin{equation}
\int_{-1}^1
    dx\,P_m(x)P_n(x)=\frac{2}{2n+1}\delta_{m,n},
\end{equation}
we obtain
\begin{eqnarray}
        \langle|\beta|^2\rangle &=&\frac{1}{N}\sum_{n=0}^\infty
    \frac{|\alpha_n(t)|^2\lambda_n}{2n+1}.
\end{eqnarray}
The far-field amplitude of the radiation scattered by $N$ atoms along the direction of the
wave-vector $\mathbf{k}=k_0(\sin\theta\cos\phi,\sin\theta\sin\phi,\cos\theta)$ is
\begin{eqnarray}
E_s(\mathbf{k})&=&E_1\sum_{j=1}^N\beta_j e^{-i\mathbf{k}\cdot \mathbf{r}_j}
\end{eqnarray}
where $E_1=(dk_0^2/4\pi\epsilon_0 r)\exp(ik_0r)$. For a continuous distribution,
\begin{eqnarray}
E_s(\mathbf{k})&=&
E_1
\int_0^{2\pi}d\phi'\int_0^\pi\sin\theta'd\theta'\int_0^\infty r'^2n(r')\beta(r',\theta')
e^{-ik_0r'[\sin\theta\sin\theta'\cos(\phi-\phi')+\cos\theta\cos\theta']}dr' \nonumber\\
&=&2\pi E_1
\int_0^\pi\sin\theta'd\theta'\int_0^\infty r'^2n(r')\beta(r',\theta')J_0(k_0r'\sin\theta\sin\theta')
e^{-ik_0r'\cos\theta\cos\theta'}dr' \nonumber\\
&=&E_1\sum_{n=0}^\infty\alpha_n i^{-n}\lambda_n
P_n(\cos\theta)
\end{eqnarray}
where $J_0(x)$ is the zero-order Bessel function and we used the integral
\begin{eqnarray}
\int_0^\pi \sin\theta'P_n(\cos\theta')J_0(k_0r'\sin\theta\sin\theta')
e^{-ik_0r'\cos\theta\cos\theta'}d\theta'=2i^{-n}j_n(k_0r')P_n(\cos\theta).
\end{eqnarray}
The angular distribution of the power scattered by $N$ atoms is
\begin{eqnarray}\label{Int}
    \frac{dP}{d\Omega}&=&\frac{c\epsilon_0}{2} |E_s(\mathbf{k})|r^2=\frac{P_1}{4\pi}
    \left|\sum_{j=1}^N\beta_j e^{-i\mathbf{k}\cdot \mathbf{r}_j}\right|^2\nonumber\\
    &=&\frac{P_1}{4\pi}\left\{\sum_j|\beta_j|^2+\sum_j\sum_{m\neq j}\beta_j\beta_m^*e^{-i\mathbf{k}\cdot(\mathbf{r}_j-\mathbf{r}_m)}\right\}
\end{eqnarray}
where $P_1=ck_0^4d^2/(32\pi^2\epsilon_0)$. The total scattered power is obtained by integrating over the solid angle, giving
\begin{equation}
P=P_1\sum_{j=1}^N\sum_{m=1}^N\beta_j\beta_m^*\frac{\sin(k_0|\mathbf{r}_j-\mathbf{r}_m|)}{k_0|\mathbf{r}_j-\mathbf{r}_m|}.\label{P:discrete}
\end{equation}
For a continuous distribution,
\begin{equation}\label{Int3}
    \frac{dP}{d\Omega}=\frac{P_1}{4\pi}\left\{N\langle|\beta|^2\rangle+\left|\sum_{n=0}^\infty\alpha_n i^{-n}\lambda_n P_n(\cos\theta)    
    \right|^2\right\}
\end{equation}
By integrating over the solid angle $4\pi$, the total scattered power is
\begin{equation}
P=P_1
\sum_{n=0}^\infty\frac{|\alpha_n|^2\lambda_n(1+\lambda_n)}{2n+1}.
\end{equation}

\section{Specific radial distribution}\label{sec:4}

We consider three different spherical distributions for which exact analytic expressions can be obtained. These include a sphere with uniform, parabolic and Gaussian profile.

\subsection{Uniform sphere \cite{Svidzinsky2008,Svidzinsky2010}}\label{uniform}

For an uniform sphere of radius $R$ and density $n(r)=N/V$ where $V=(4\pi/3)R^3$ and $0<r<R$,
\begin{equation}
F_n(r)=\frac{3N}{R^3}\left\{h^{(1)}_{n}(k_0 r)\int_0^r dr' r'^2j^2_{n}(k_0 r')
    +
    j_{n}(k_0 r)\int_r^R dr' r'^2j_n(k_0r')h^{(1)}_{n}(k_0 r')\right\}
\end{equation}
Taking $r=R$ and defining $k_0R=\sigma$, we obtain
\begin{equation}
F_n(\sigma)=\frac{3N}{\sigma^3}h^{(1)}_{n}(\sigma)\int_0^\sigma dx x^2j^2_{n}(x).
\end{equation}
Since
\begin{eqnarray}
\int x^2 j^2_n(x)dx &=& \frac{x^3}{2}\left\{j_n^2(x)-j_{n-1}(x)j_{n+1}(x)\right\}
\end{eqnarray}
we obtain
\begin{equation}
F_n(\sigma)=h^{(1)}_{n}(\sigma)\lambda_n
\end{equation}
where
\begin{eqnarray}
\lambda_n&=& \frac{3N}{2}\left\{j^2_n(\sigma)-j_{n-1}(\sigma)j_{n+1}(\sigma)\right\}\label{lambda:n}
\end{eqnarray}
is the collective decay rate of the mode $n$.
By inserting these expressions in Eq.(\ref{Ebeta3}) with $r=R$ we obtain, for $j_n(\sigma)\neq 0$,
\begin{eqnarray}
    \dot\alpha_{n}-\Gamma\left\{i(\delta-\omega_n)-\frac{1}{2}(1+\lambda_n)\right\}\alpha_n=-i\frac{\Omega_0}{2}(2n+1)i^n
    \label{alpha:2}
\end{eqnarray}
where $\delta=\Delta_0/\Gamma$ and  $\omega_n=[y_n(\sigma)/j_n(\sigma)]\lambda_n/2$ is the collective Lamb shift of the mode $n$.
Equation (\ref{alpha:2}) can be straightforwardly integrated and, once inserted in
Eq. (\ref{betaexp}), leads to the following expression for the excitation amplitude
\begin{equation}\label{betasol:1}
    \beta(r,\theta,t)=\frac{\Omega_0}{\Gamma}\sum_{n=0}^\infty\frac{i^{n}(2n+1)j_n(k_0 r)P_{n}(\cos\theta)}
    {2(\delta-\omega_n)+i(1+\lambda_n)}
    \left[1-e^{i(\delta-\omega_n)\Gamma t-(1+\lambda_n)\Gamma t/2}\right]
\end{equation}
If the pump is switched off after the steady-state is reached (taken as the time $t=0$),
\begin{eqnarray}
    \beta^{\mathrm{(free)}}(r,\theta,t)=\frac{\Omega_0}{\Gamma}\sum_{n=0}^\infty\frac{i^{n}(2n+1)j_n(k_0 r)P_{n}(\cos\theta)}
    {2(\delta-\omega_n)+i(1+\lambda_n)}e^{i(\delta-\omega_n)\Gamma t-(1+\lambda_n)\Gamma t/2}\label{betafin:1}.
\end{eqnarray}
Then
\begin{equation}
\langle|\beta^{\mathrm{(free)}}|^2\rangle=\frac{\Omega_0^2}{N\Gamma^2}\sum_{n=0}^\infty\frac{(2n+1)\lambda_n}
    {4(\delta-\omega_n)^2+(1+\lambda_n)^2}e^{-(1+\lambda_n)\Gamma t},\label{ave:beta}
\end{equation}
\begin{equation}
    \frac{dP}{d\Omega}=\frac{P_1}{4\pi}\left\{
    \sum_{n=0}^\infty\frac{(2n+1)\lambda_n e^{-(1+\lambda_n)\Gamma t}}
    {4(\delta-\omega_n)^2+(1+\lambda_n)^2} 
    +\left|\sum_{n=0}^\infty
    \frac{(2n+1)\lambda_n P_n(\cos\theta)}
    {2(\delta-\omega_n)^2+i(1+\lambda_n)} e^{-i\omega_n\Gamma t-(1+\lambda_n)\Gamma t/2}
    \right|^2\right\}\label{dP:dOmega}
\end{equation}
and
\begin{equation}
P(t)=P_1 \sum_{n=0}^\infty\frac{(2n+1)\lambda_n(1+\lambda_n) e^{-(1+\lambda_n)\Gamma t}   }
    {4(\delta-\omega_n)^2+(1+\lambda_n)^2}.\label{P}
\end{equation}
We observe that this solution does not describe the subradiant decay after the laser is cut off, since every mode has a decay rate $(1+\lambda_n)\Gamma>\Gamma$ i.e. larger than the single-atom decay. The MF model is unable to describe subradiance, experimentally observed in \cite{Araujo2016} and theoretically discussed in \cite{Bienaime2012}: single photon subradiance arises from the anti-symmetric states of $N$ atoms, in which only a single excitation among $N$ is present \citep{Dicke1954,Scully2015}. Hence, it can be described only by the discrete model of Eq.(\ref{betaj}). Conversely, single-photon superradiance can be well described by the MF model, as it will discussed in the following.

\begin{figure}[t]
       \centerline{\scalebox{0.3}{\includegraphics{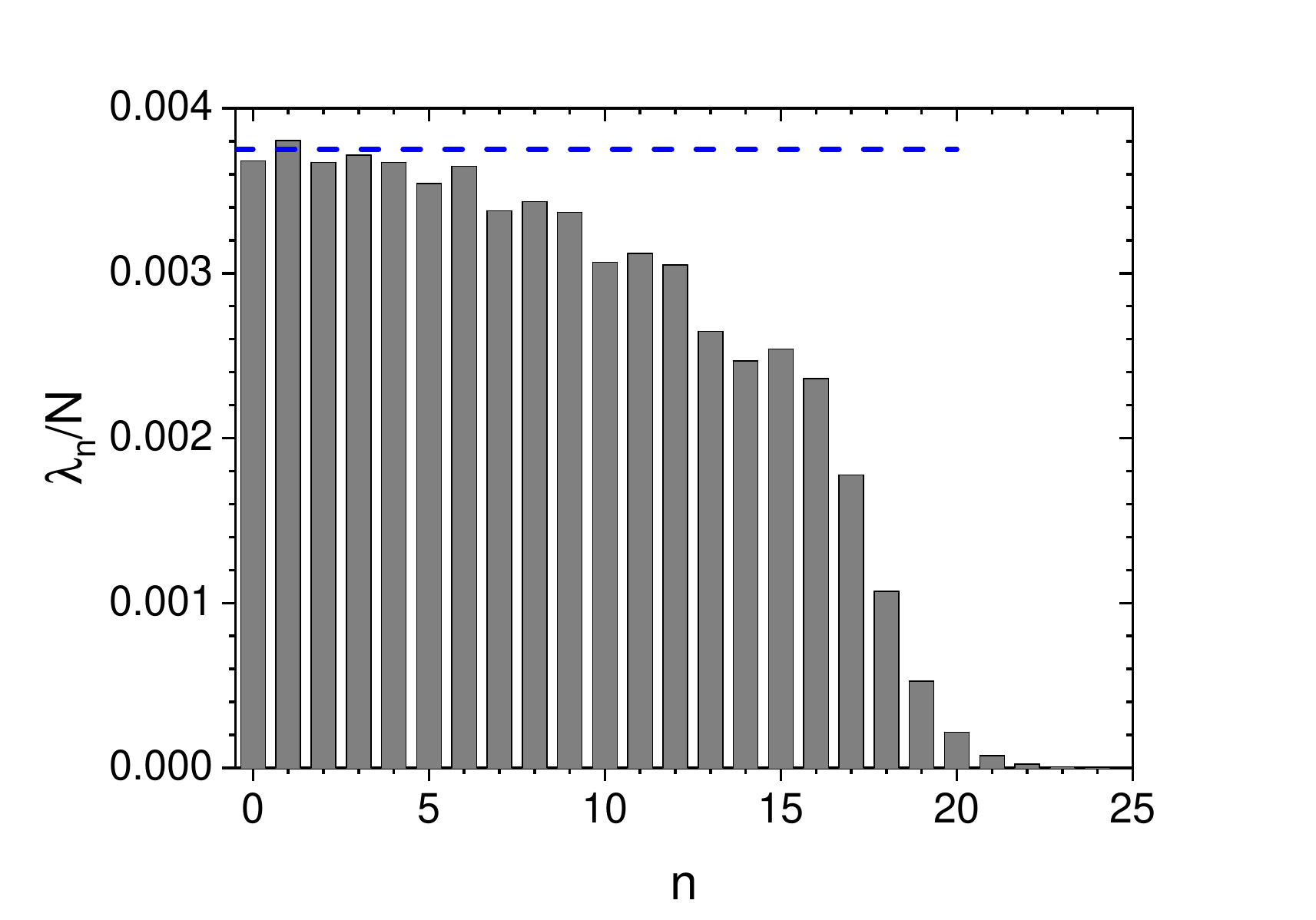}}}
        \caption{$\lambda_n/N$ for an uniform sphere with $\sigma=20$. The dashed blue line is the value $\lambda_N/N=3/2\sigma^2$.}
        \label{Fig1}
    \end{figure}

\begin{figure}
      \centerline{\scalebox{0.3}{\includegraphics{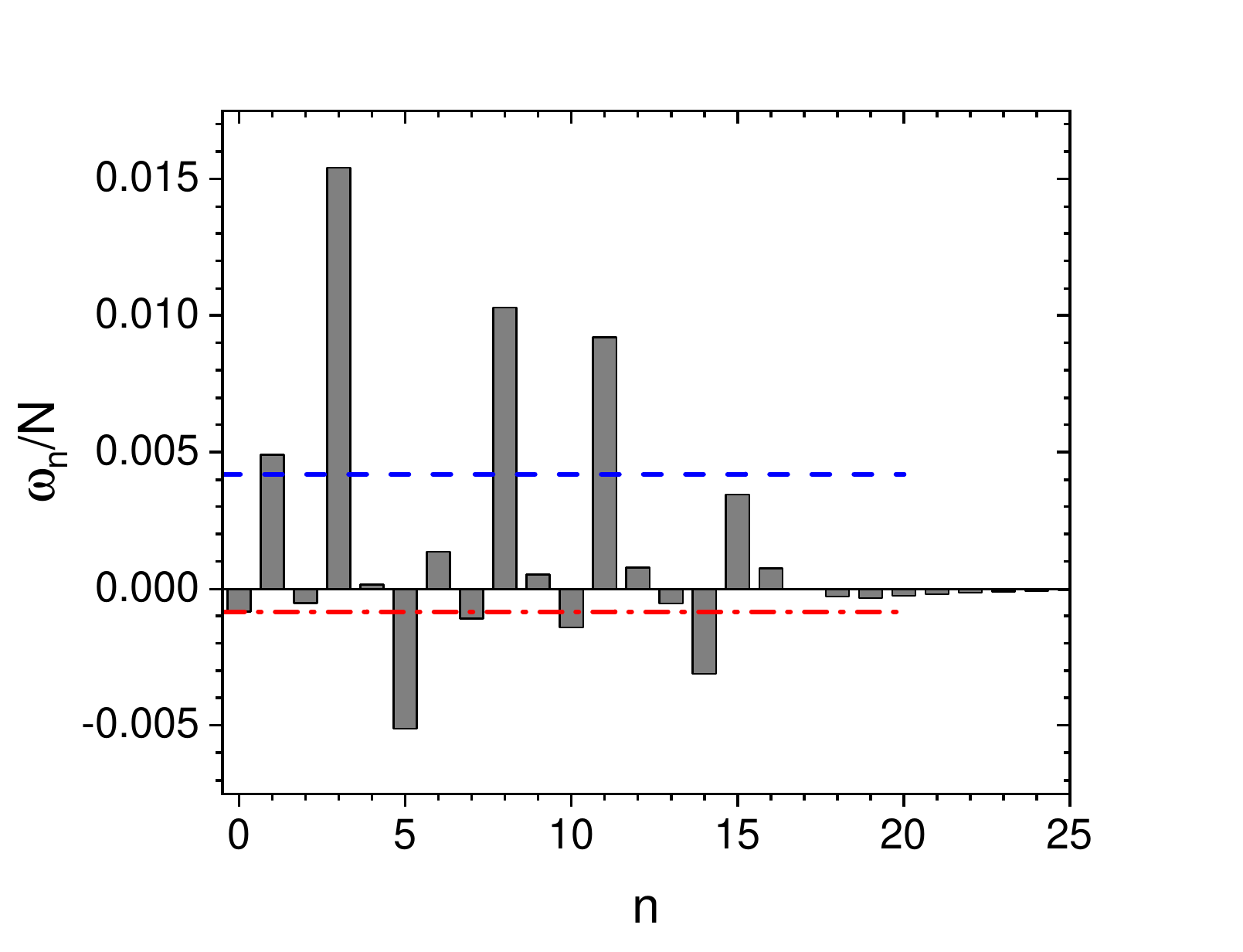}}}
        \caption{$\omega_n/N$ for an uniform sphere with $\sigma=20$. The dashed blue line is $\omega_N/N=3/4\sigma^2\tan\sigma$, the dash-dotted red line is the value $\omega_N/N=-3/4\sigma^2\cot\sigma$.}
        \label{Fig2}
\end{figure}

For a small cloud, with $\sigma\ll 1$, only the term $n=0$, with $\lambda_0\approx N$, decays fast (Dicke superradiance \cite{Dicke1954}), while all the other terms with $n\ge 1$ are suppressed by a factor $\sigma^{2n}$. The collective shift is $\omega_0\sim -N/2\sigma$.
The case of a large cloud is illustrated by 
Fig.\ref{Fig1} and \ref{Fig2}, showing $\lambda_n/N$ and $\omega_n/N$ for $\sigma=20$, as obtained from Eq.(\ref{lambda:n}). We observe that
for $\sigma\gg 1$ and $n<\sigma$, $\lambda_n\approx 3N/2\sigma^2\equiv\lambda_N$ (dashed blue line in Fig.\ref{Fig1}) is almost independent on $n$
and drops to zero for $n>\sigma$, approximately as
\begin{equation}
\lambda_n\approx \frac{3\pi N}{4n^2n!^2}\left(\frac{\sigma^2}{4}\right)^n.
\end{equation}
The collective Lamb shift $\omega_n$ in the limit $\sigma\gg 1$ and $n<\sigma$ is approximately
   $\omega_n\sim -(\lambda_N/2)\cot(\sigma-n\pi/2)\sim (3N/4\sigma^2)\{\tan\sigma,-\cot\sigma\}$, where the first value is for $n$ odd and the second for $n$ even (dashed blue line and dash-dotted red line in Fig.\ref{Fig2}, respectively).
We observe that $\omega_n$ changes sign with $n$ and, with the exception for the values of $\sigma$ where $\tan\sigma$ or $\cot\sigma$ are large, it averages to zero and gives a negligible contribution. For large detuning, $\delta\gg 1$, it can be neglected.

\begin{figure}
      \centerline{\scalebox{0.4}{\includegraphics{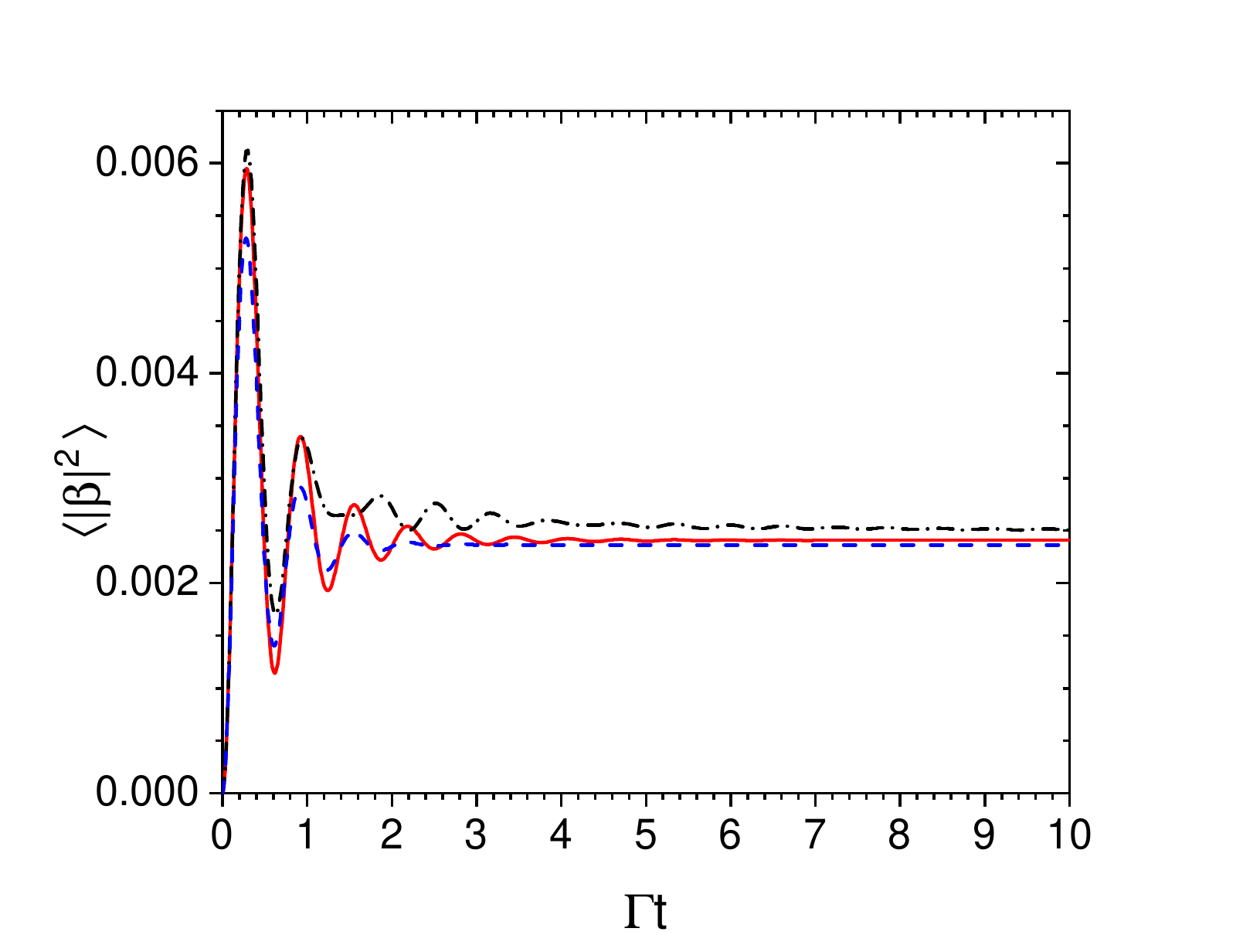}}}
        \caption{$\langle|\beta(t)|^2\rangle$ (in units of $(\Omega_0/\Gamma)^2$) vs $\Gamma t$ for $\delta=10$ and an uniform sphere with $\sigma=20$ and $N=10^3$, from the analytical MF solution (continuous red line),  from the numerical solution of the discrete equations (\ref{betaj}) (dash-dot black line) and from the Timed-Dicke approximated solution, (\ref{TD}) (dash blue line).}
        \label{Fig3}
\end{figure}

Fig.\ref{Fig3} shows the average excitation probability $\langle|\beta(t)|^2\rangle$ vs $\Gamma t$ for $\delta=10$, $\sigma=20$ and $N=10^3$: the continuous red line is the MF solution, obtained from Eq.(\ref{betasol:1}), whereas the dash black line is the numerical solution of Eqs.(\ref{betaj}). The Timed-Dicke approximated solution \cite{Courteille2010,Bienaime2010,Manassah2012a,Manassah2012} can be obtained by assuming $\lambda_n\approx\lambda_N$, giving
\begin{equation}
\langle|\beta(t)|^2\rangle=\frac{1}{N\Gamma^2}\frac{\Omega_0^2}
    {4\delta^2+(1+\lambda_N)^2}\left|1-e^{i\delta\Gamma t-(1+\lambda_N)\Gamma t/2}\right|^2.\label{TD}
\end{equation}
This solution, reported in Fig.\ref{Fig3} by the dashed blue line, is in good agreement with the exact solution, confirming that the driving laser brings the atoms into a state well described by the Timed-Dicke approximation, where the remaining subradiant part is only a small fraction of it.

\begin{figure}
      \centerline{\scalebox{0.4}{\includegraphics{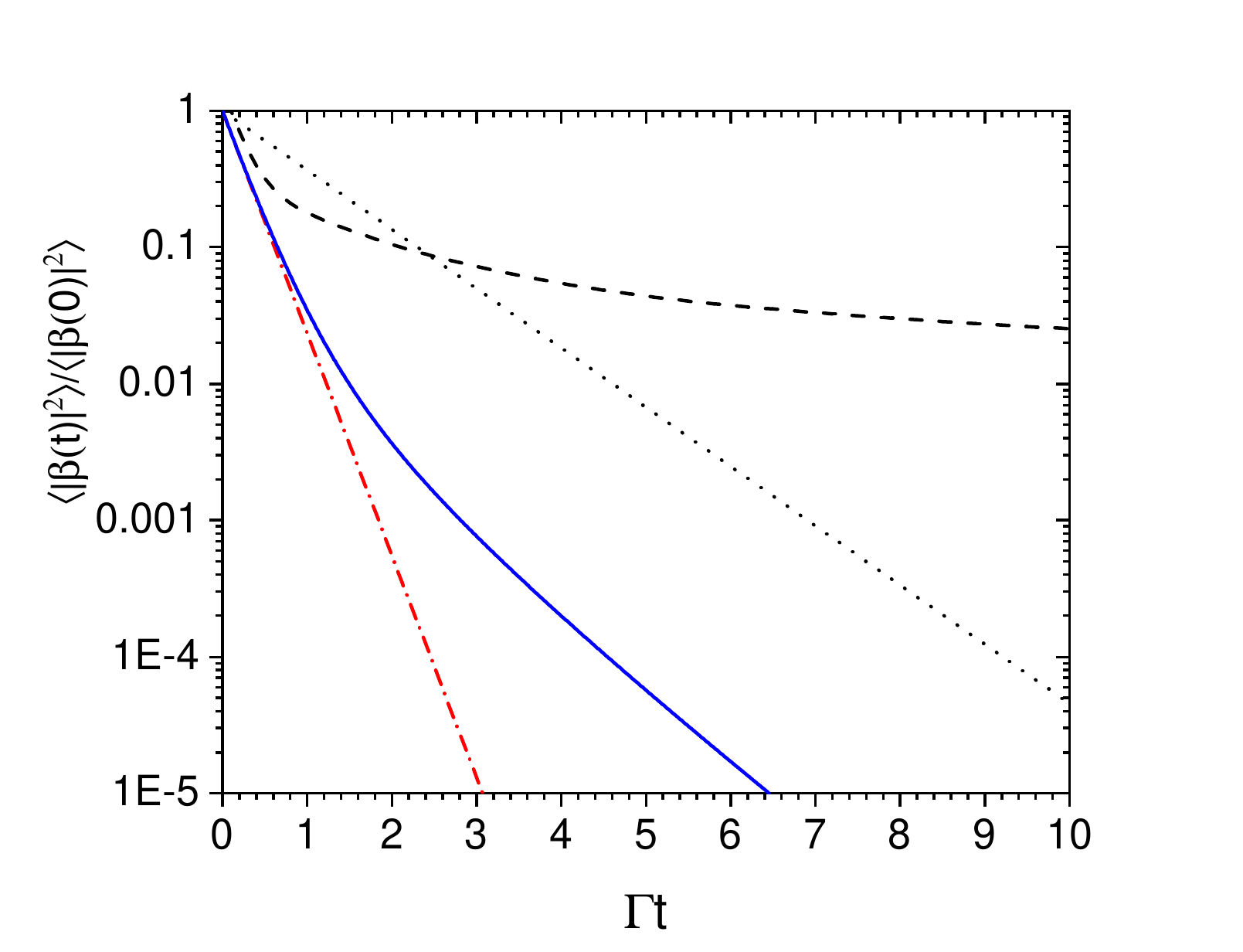}}}
        \caption{$\langle|\beta^{\mathrm{(free)}}(t)|^2\rangle/\langle|\beta^{\mathrm{(free)}}(0)|^2\rangle$ vs $\Gamma t$ for $\delta=10$ and an uniform sphere with $\sigma=20$ and $N=10^3$, from the analytical MF solution (continuous blue line) and from the numerical solution of the discrete equations (\ref{betaj}) (dashed black line). The dashed-dotted red line is the Timed-Dicke approximation, $\exp(-\lambda_N\Gamma t)$ and the dotted black line is the single atom decay $\exp(-\Gamma t)$.}
        \label{Fig4}
\end{figure}

When the laser is cut off, at short times the decay is superradiant, with $\lambda_n\approx \lambda_N$ and 
   \begin{eqnarray}
    \beta^{\mathrm{(free)}}(r,\theta,t)&\approx &\frac{(\Omega_0/\Gamma)}{2\delta+i(1+\lambda_N)}e^{i\delta\Gamma t-(1+\lambda_N)\Gamma t/2}
    \sum_{n=0}^\infty i^{n}(2n+1)j_n(k_0 r)P_{n}(\cos\theta)\nonumber\\
    &=&\frac{(\Omega_0/\Gamma)}{2\delta+i(1+\lambda_N)}
    e^{ik_0r\cos\theta+i\delta\Gamma t-(1+\lambda_N)\Gamma t/2}
    \label{beta:TD}.
\end{eqnarray}
Fig.\ref{Fig4} shows $\langle|\beta^{\mathrm{(free)}}(t)|^2\rangle/\langle|\beta^{\mathrm{(free)}}(0)|^2\rangle$ vs $\Gamma t$ in semi-log scale for the same parameters of Fig.\ref{Fig3}, after the laser is cut off. The continuous blue line is the MF solution, Eq. (\ref{ave:beta}), the dashed black line is the numerical solution of Eqs.(\ref{betaj}), the dashed-dotted red line is the Timed-Dicke superradiant decay $\exp(-\lambda_N\Gamma t)$ and the dotted black line is the single-atom decay $\exp(-\Gamma t)$. We observe that the MF solution initially follows the fast superradiant decay as $\exp(-\lambda_N\Gamma t)$ and later the single-atom decay $\exp(-\Gamma t)$. Instead, the discrete solution shows a subradiant decay, slower than the single-atom decay. This behavior is peculiar of the discrete system and can not be caught by the MF model. 

   
 \subsection{Parabolic profile}
 
\begin{figure}[t]
       \centerline{\scalebox{0.3}{\includegraphics{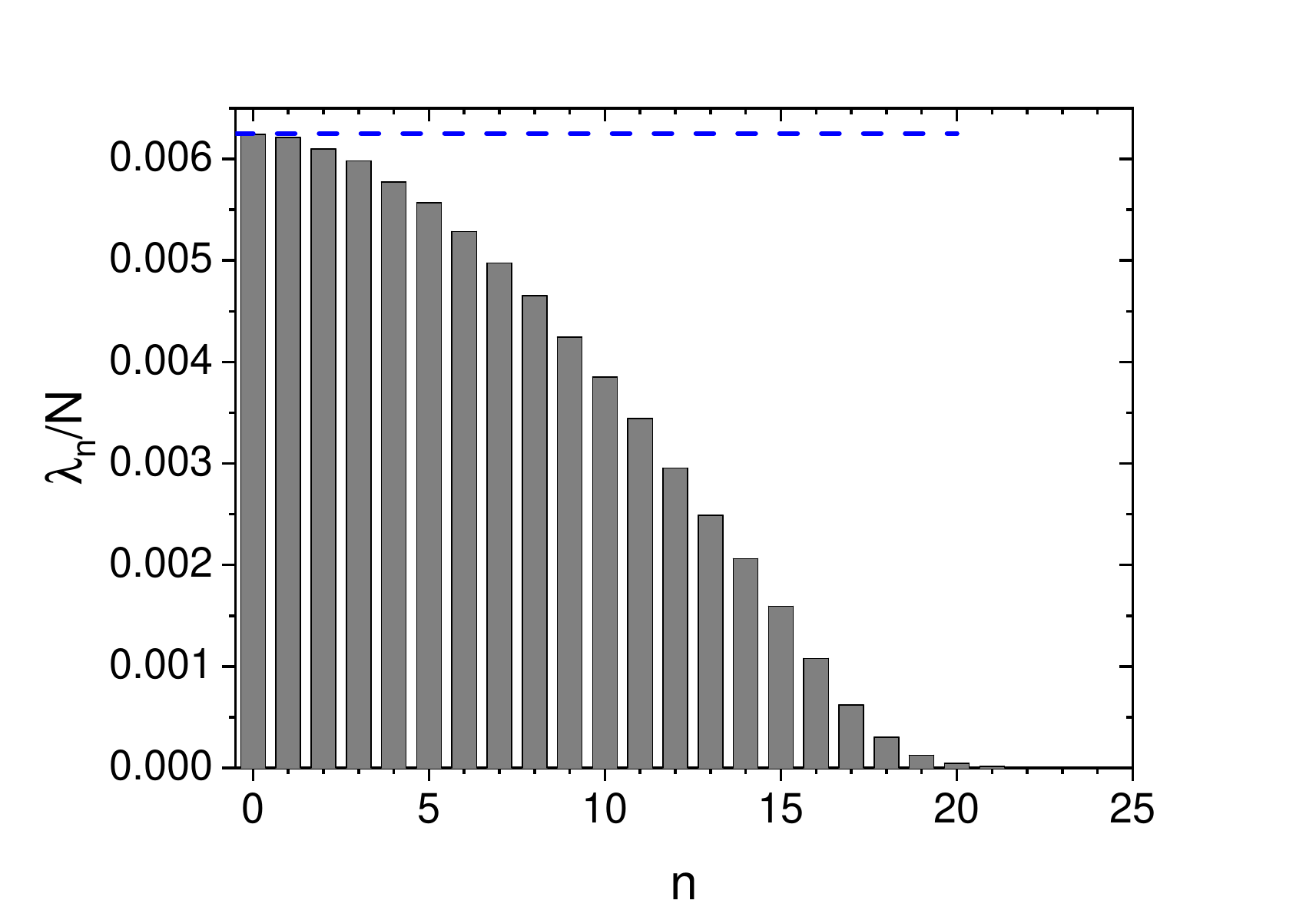}}}
        \caption{$\lambda_n/N$ for a sphere with parabolic profile, with $\sigma=20$. The dashed blue line is the value $\lambda_N/N=5/2\sigma^2$.}
        \label{Fig5}
    \end{figure}

 An other case which can be solved analytically is a sphere with a parabolic profile,  with radial density
 $n(r)=(15 N/8\pi R^3)(1-r^2/R^2)$ and  $0<r<R$. In this case we obtain
 \begin{eqnarray}
 \lambda_n&=&\frac{15 N}{2}\left\{
 \frac{1}{3}j_n^2(\sigma)-\frac{1}{2}j_{n+1}(\sigma)j_{n-1}(\sigma)-\frac{1}{6}j_{n-1}^2(\sigma)\right.\nonumber\\
 &+&\left.\frac{1}{3\sigma}\left(n+\frac{3}{2}\right)j_{n-1}(\sigma)j_{n}(\sigma)\right.\nonumber\\
 &-&\left.\frac{1}{3\sigma^2}\left(n+\frac{3}{2}\right)
 \left[\left(n+\frac{1}{2}\right)j_n^2(\sigma)-\left(n-\frac{1}{2}\right)j_{n+1}(\sigma)j_{n-1}(\sigma)\right]
 \right\}\label{lambda:par}.
 \end{eqnarray}
 where $\sigma=k_0R$. The other expressions, obtained from the uniform sphere in Sec.\ref{uniform}, remain valid. Fig.\ref{Fig5}
 shows $\lambda_n/N$ for $\sigma=20$, as obtained from Eq.(\ref{lambda:par}). For $\sigma\gg 1$ and $n\ll\sigma$, 
 $\lambda_n\approx (5N/2\sigma^2)$. 
 
 \subsection{Gaussian profile} 
 
\begin{figure}[t]
       \centerline{\scalebox{0.3}{\includegraphics{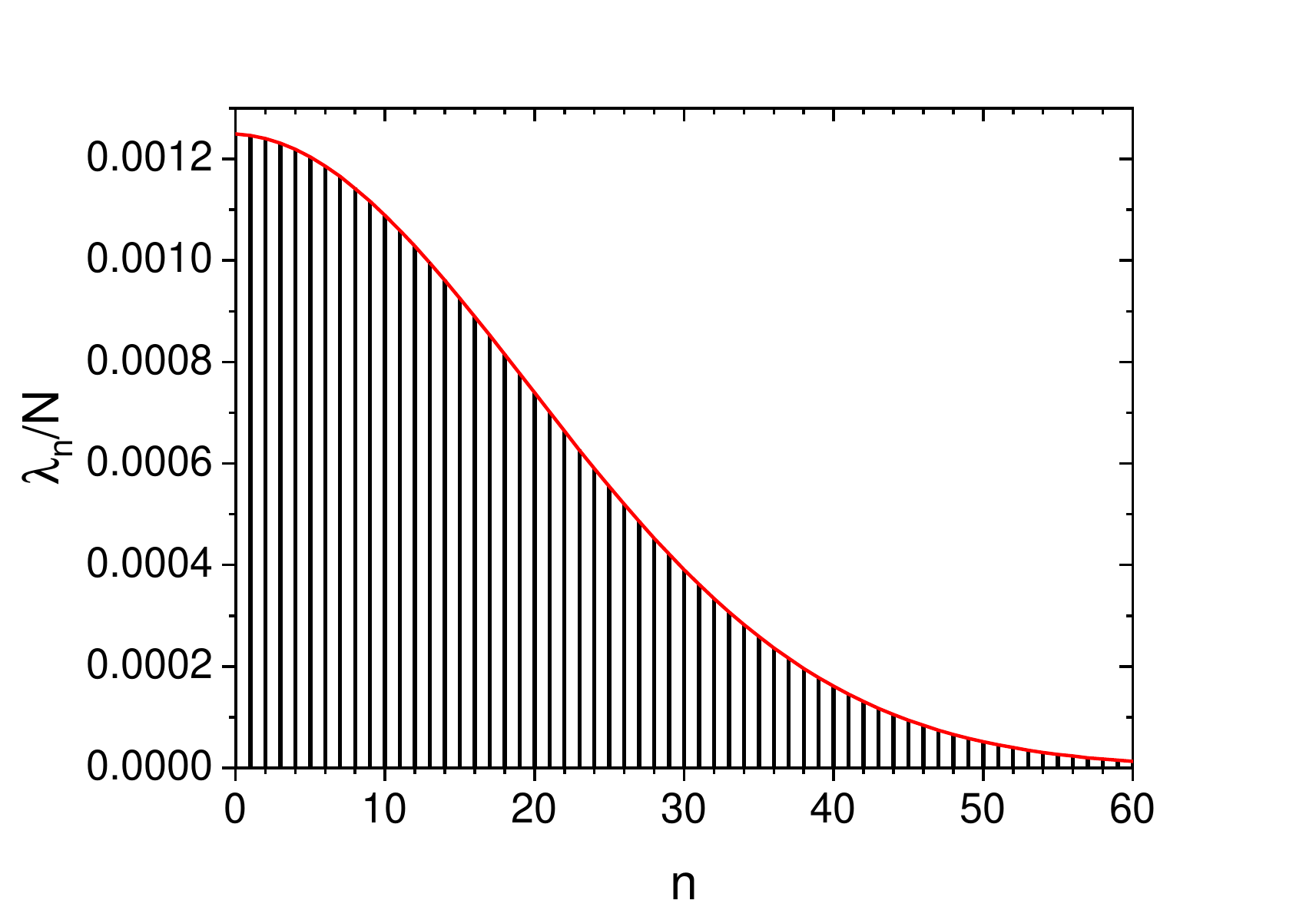}}}
        \caption{$\lambda_n/N$ vs $n$ for a sphere with Gaussian profile, with $\sigma=20$, for the exact discrete expression (\ref{lambda_n:gauss}) and its continuous approximation (\ref{lambda:gauss}) (red continuous line).}
        \label{Fig6}
    \end{figure}

 For a Gaussian profile, with density $n(r)=[N/(2\pi)^{3/2}\sigma_R^3]\exp(-r^2/2\sigma_R^2)$, we obtain \cite{Bachelard2011}
 \begin{equation}
 \lambda_n=N\sqrt{\frac{\pi}{2\sigma}}e^{-\sigma^2}I_{n+1/2}(\sigma^2)\label{lambda_n:gauss}
 \end{equation}
 where $\sigma=k_0\sigma_R$ and $I_n(x)$ is the $n$th-order modified Bessel function. Taking the limit $r\rightarrow\infty$ in Eq.(\ref{Fn}),
 we obtain the same equation (\ref{alpha:2}) for $\alpha_n(t)$ and the same expression (\ref{ave:beta}) as for the uniform sphere, where the collective shift $\omega_n=(\lambda_n/2)\lim_{r\rightarrow\infty}\{y_n(k_0r)/j_n(k_0r)\}$ may be neglected. For $\sigma$ large, all the modes up to $n\sim \sigma$ are significant and
 \begin{equation}
 \lambda_n\approx \frac{N}{2\sigma^2}e^{-(n+1/2)^2/2\sigma^2}\label{lambda:gauss}.
 \end{equation}
 The spectrum can be treated as a continuum, with $\lambda_n\approx \lambda(\eta)=(N/2\sigma^2)\exp(-\eta^2/2\sigma^2)$ (where $\eta=n+1/2$).
 Fig.\ref{Fig6} shows the discrete values $\lambda_n/N$ vs $n$ for $\sigma=20$ from Eq.(\ref{lambda_n:gauss}) (columns) and its continuous approximation (\ref{lambda:gauss}) (red continuous line). 
  
\begin{figure}
      \centerline{\scalebox{0.4}{\includegraphics{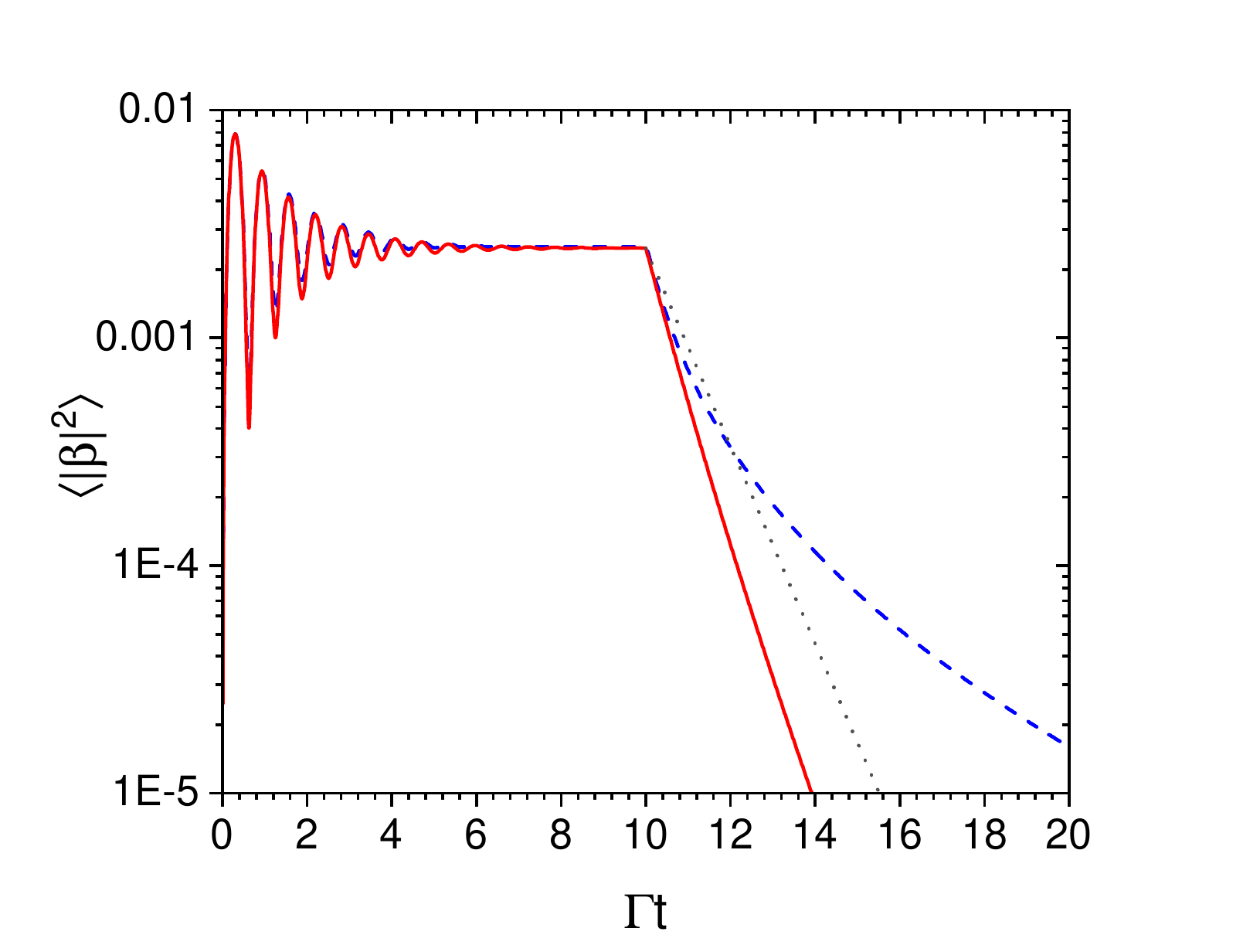}}}
        \caption{$\langle|\beta(t)|^2\rangle$ (in units of $(\Omega_0/\Gamma)^2$) vs $\Gamma t$ for $\delta=10$ and a Gaussian sphere with $\sigma=20$ and $N=10^3$, from the analytical MF solution (continuous red line) and from the numerical solution of the discrete equations (\ref{betaj}) (dashed blue line). The dotted black line is the single-atom decay, as $\exp(-\Gamma t)$.}
        \label{Fig7}
\end{figure}

 Then, the sum in Eq.(\ref{ave:beta}) can be approximated by an integral, $\sum_{n=0}^\infty(2n+1)\rightarrow 2\int_0^\infty \eta d\eta$ to get
\begin{eqnarray}
\langle|\beta^{\mathrm{(free)}}|^2\rangle&=&\frac{2\Omega_0^2}{N\Gamma^2}\int_{0}^\infty\frac{\eta\lambda(\eta)}
    {4\delta^2+(1+\lambda(\eta))^2}e^{-(1+\lambda(\eta))\Gamma t}d\eta\nonumber\\
    &=& \left(\frac{\Omega_0}{\Gamma}\right)^2\frac{2\sigma^2}{N}\int_{0}^{N/2\sigma^2}\frac{dx}
    {4\delta^2+(1+x)^2}e^{-(1+x)\Gamma t},\label{ave:beta:gauss}
\end{eqnarray}
where we have set $x=\lambda(\eta)$. In the limit $\delta\gg 1$,
\begin{eqnarray}
\langle|\beta^{\mathrm{(free)}}|^2\rangle
&=&\left(\frac{\Omega_0}{2\delta\Gamma}\right)^2
\frac{e^{-\Gamma t}}{\Gamma_{sr}t}\left(1-e^{-\Gamma_{sr}t}\right),
\end{eqnarray}
where $\Gamma_{sr}=(N/2\sigma^2)\Gamma$ is the superradiant decay rate. Instead, for $\delta=0$
\begin{eqnarray}
\langle|\beta^{\mathrm{(free)}}|^2\rangle
&=&\left(\frac{\Omega_0}{\Gamma}\right)^2\frac{2\sigma^2}{N}\Gamma t\{\gamma(-1,(\Gamma+\Gamma_{sr})t)-\gamma(-1,\Gamma t)\},
\end{eqnarray}
where $\gamma(a,x)=\int_0^x e^{-u}u^{a-1}du$ is the lower incomplete gamma function. For large times, it can be approximated by
\begin{eqnarray}
\langle|\beta^{\mathrm{(free)}}|^2\rangle
&\approx &\left(\frac{\Omega_0}{\Gamma}\right)^2
\frac{e^{-\Gamma t}}{\Gamma_{sr}t}\left[1-\left(\frac{\Gamma_{sr}}{\Gamma+\Gamma_{sr}}\right)^2e^{-\Gamma_{sr}t}\right],
\end{eqnarray}
Hence, the decay of the excitation is not exponential, neither in the superradiant regime: at short times the decay rate is $\Gamma_{sr}$ and at later times the excitation decays as $\exp(-\Gamma t)/(\Gamma_{sr}t)$, before the slower subradiant decay takes place at time larger than $1/\Gamma$. Fig.\ref{Fig7} shows $\langle|\beta(t)|^2\rangle$ vs $\Gamma t$ for $\delta=10$ and a Gaussian sphere with $\sigma=20$ and $N=10^3$, from the analytical MF solution (continuous red line) and from the numerical solution of the discrete equations (\ref{betaj}) (dashed blue line).
We observe a good agreement between the MF and the discrete models as long as the laser is on. Just after the laser is cut, the two solutions show that the excitation decays superradiantly, with a rate $\Gamma_{sr}$, but at later times the exact discrete model shows that the decay is subradiant, with a rate less than the single-atom value $\Gamma$ (shown by the dotted black line in Fig.\ref{Fig7}).

\begin{figure}
      \centerline{\scalebox{0.4}{\includegraphics{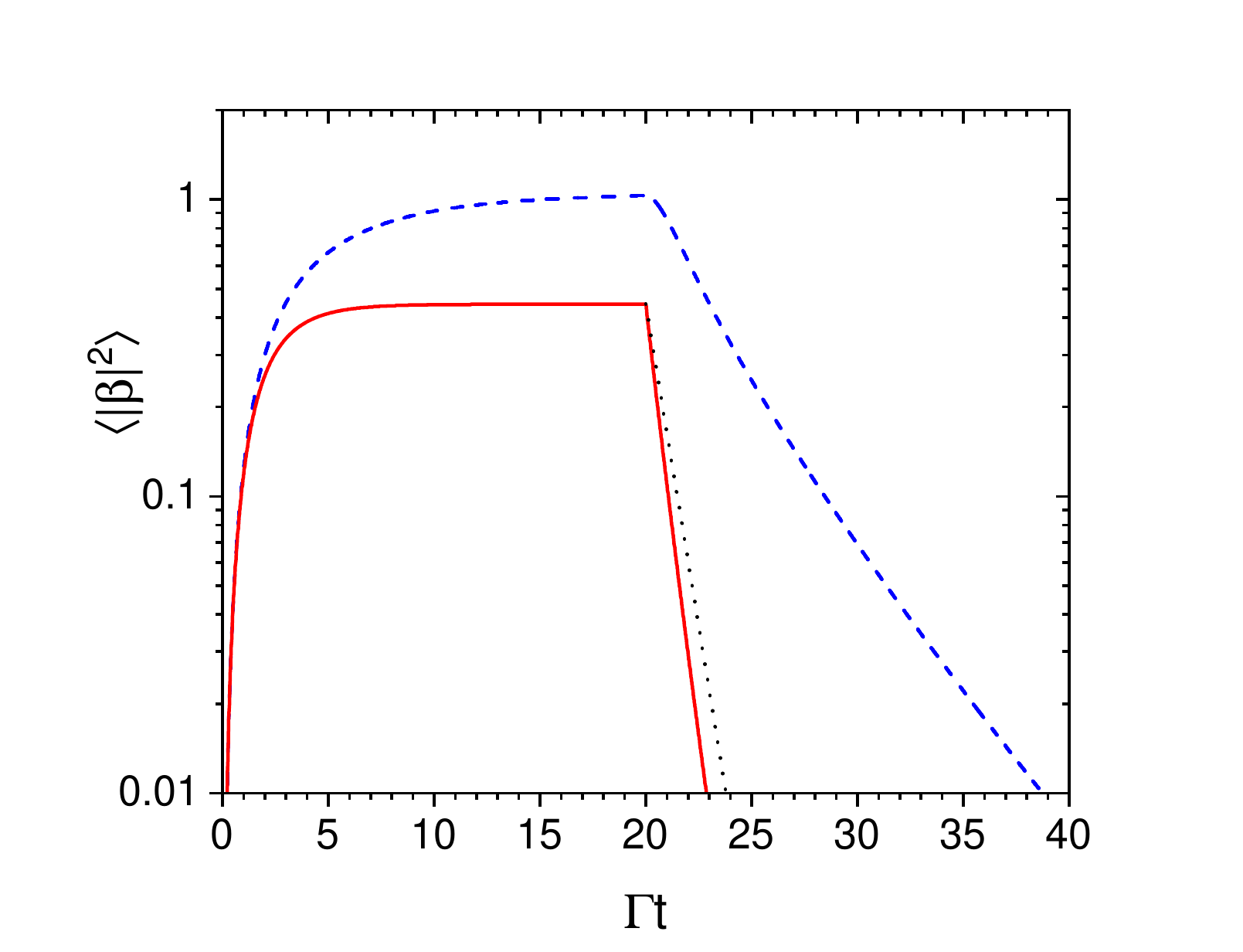}}}
        \caption{$\langle|\beta(t)|^2\rangle$ (in units of $(\Omega_0/\Gamma)^2$) vs $\Gamma t$ for $\delta=0$ and a Gaussian sphere with $\sigma=20$ and $N=10^3$, from the analytical MF solution (continuous red line) and from the numerical solution of the discrete equations (\ref{betaj}) (dashed blue line). The dotted black line is the single-atom decay, as $\exp(-\Gamma t)$.}
        \label{Fig8}
\end{figure}

\begin{figure}
      \centerline{\scalebox{0.4}{\includegraphics{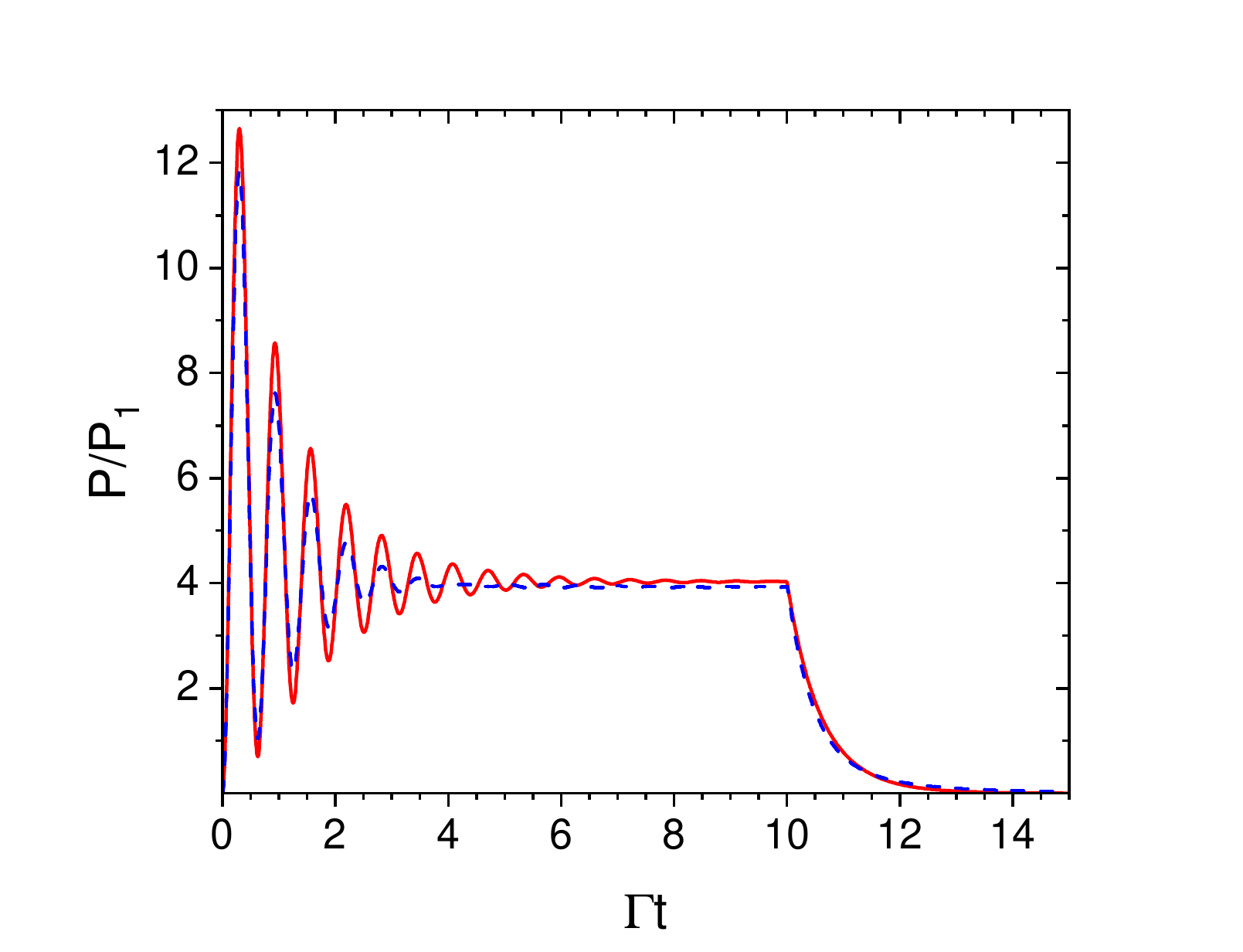}}}
        \caption{$P/P_1$ vs $\Gamma t$ for $\delta=10$ and a Gaussian sphere with $\sigma=20$ and $N=10^3$, from the analytical MF solution (continuous red line) and from the numerical solution of the discrete equations (\ref{P:discrete}) (dashed blue line).}
        \label{Fig9}
\end{figure}

Figure \ref{Fig8} shows $\langle|\beta(t)|^2\rangle$ vs time for the same case of Fig.\ref{Fig7}, except that now $\delta=0$. In this case the MF solution (red continuous line) does not reproduce well the exact discrete solution (dashed blue line), neither when the laser is on. This confirms that the MF solution does not describe the multiple-scattering regime (and hence the diffusion regime),  characterized by a large optical thickness $b=b_0/(1+4\delta^2)$ (where $b_0=3N/\sigma^2$ is the resonant optical thickness). In the case of Fig.\ref{Fig8}, $\delta=0$ and $b=b_0\sim 7.5$, whereas in the case of Fig.\ref{Fig7} $\delta=10$ and $b\ll 1$. In the MF model the interaction is coherent and dominated by collective modes: in order to describe the diffusive dynamics, where the particles scatter many photons in a mean-free path, the model must include granularity and fluctuations, which are missed assuming a smooth, continuous density distribution.
Finally Fig. \ref{Fig9} shows the total scattered power vs time $P_s(t)$ (in units of the single-atom value $P_1$), calculated from the MF model, Eq.(\ref{P}), (continuous red line) and for the exact discrete model, Eq.(\ref{P:discrete}), (dashed blue line). The parameters are those of Fig.\ref{Fig7}. The MF solution describes rather well the exact behavior, also if the transient oscillations are more strongly damped in the exact solution. Just after the laser is cut, the decay rate is superradiant, with a rate $\Gamma_{sr}=N\Gamma/2\sigma^2$ proportional to the resonant optical thickness. Subradiant decay occurs at later times, after the power has decreased by several orders of magnitude.

\section{Conclusions}\label{sec:5}

The aim of this paper has been to provide an analytical description of the cooperative light scattering by an ensemble of atoms driven by an uniform laser beam. We have compared the mean-field (MF) model, where a continuous atomic distribution is assumed, to the numerical results from the discrete coupled dipoles model. The MF model describes a coherent interaction between the atoms, neglecting multiple scattering and diffusion effects due to the random walk of the photon within a mean-free pass distance. For these reasons, the validity of the MF model is limited to a regime with small optical thickness $b=b_0/(1+4\delta^2)\ll 1$, but still cooperative when $b_0\gg 1$ and $\delta\gg 1$. In this regime the MF model gives a rather accurate description of the atomic excitation and of the scattered light intensity when the laser is on, but is unable to describe the subradiant decay after the laser is cut off. This suggests that subradiance is intrinsically related to the discreetness of the system and to the anti-symmetric properties of the single-excitation $N$-atomic states. Contrarily to previous works, we do not assume an initial preparation of the atoms in a superposition of states with a single excitation (the so-called Dicke states), but the excitation is provided by a classical uniform laser. The atomic system reaches a stationary state which is dominated by the Timed-Dicke symmetric state. When the laser is cut, the early decay is superradiant, with a rate $\Gamma_{sr}\sim N\Gamma/(k_0 R)^2$, where $R$ is the size of the atomic cloud. The MF solution can be expressed in terms of collective modes whose features depend on the atomic distribution. We discussed the cases of uniform, parabolic and Gaussian spherical distribution. When the cloud's size is smaller than an optical wavelength, a single mode with decay rate $N\Gamma$ will dominate, whereas for an extended cloud many modes are present, up to a number $n\sim k_0 R$: the fastest modes are those with a decay rate proportional to the resonant optical thickness $b_0$, down to the slower ones with decay rate $\Gamma$. So, the last surviving modes when the laser is off are those with a single-atom decay rate. In this sense, the subradiant component of the excited state is lost in a MF description.

\bibliography{Bibliography}

\end{document}